\documentclass[12pt,thmsa]{article}
\usepackage{sw20aip}

\input{tcilatex}
\begin{document}

\begin{titlepage}

\begin{center}
{\Large  \bf The relation of thermal fluctuation and information-entropy of 
One-dimensional Rindler Oscillator}

\vspace{5mm}

\end{center}

\vspace{5 mm}

\begin{center}
{\bf Dan Yu, Yuan-Xing Gui \footnote{Corresponding author. E-mail: guiyx@dlut.edu.cn},
 and  Xin Ye}

\vspace{3mm}
{\it Department of Physics, Dalian University of Technology\\
  Dalian, Liaoning 116024, P.R. China\\}

\end{center}

\vspace{1cm}


\baselineskip=24pt
Within the framework of thermo-field-dynamics(TFD),
the information-entropies associated with the measurements of
position and momentum for one-dimensional Rindler oscillator are derived,
and the connection between its information-entropy and  thermal fluctuation is obtained. 
A conclusion is drawn that the thermal fluctuation leads to the loss of information.\\

\vspace{1cm}
{\bf PACS number(s)}: 04.90.+e, 05.40+j, 03.65.-w
\end{titlepage}
\baselineskip=24pt

\section{\textbf{\ INTRODUCTION }}

As a joint-realm of information theory and statistical physics, \smallskip
information-entropy has received a great deal of investigations$^{1-3}$.
Generally, for a thermal state, the thermal fluctuation results in losses of
information and increases of the uncertainty and information-entropy.
Therefore our interest here concerns what is the connection between the
thermal fluctuation and information-entropies.

The generalized uncertainty relation $^{4-8}$ is widely discussed. But in
these papers the ``generalized uncertainty relations'' have different
meanings. In our previous paper$^4$, the generalized uncertainty relation of
one-dimensional Rindler oscillator, which is related to the thermal nature
of the quantum state, was derived. In fact, we presented a thermal
modification to the uncertainty principle, it is the acceleration that
induces the thermal effect (Unruh effect). As is well know, a Minkowski
vacuum is equivalent to a thermal bath$^{9,10}$for a Rindler observer. For a
Rindler uniformly accelerated observer, there are not only general quantum
fluctuations but also thermal fluctuations related to his acceleration. On
the other hand, the authors$^{6,7}$ presented a gravitationally-induced
modification to the momentum and position uncertainty relation which comes
from a definition of momentum and position operators that incorporates
certain dynamics relevant at high energies and differs from the usual
definition.

In this paper, we present the information-entropies associated with the
measurements of position and momentum for one-dimensional Rindler oscillator
in the coordinate representation, and derive a relation between
information-entropies and fluctuations. Compared with other methods such as
Wigner function method$^{11,12}$, our method introduced in this paper is
different from others. Our strategy is as follows. First we calculate the
information-entropies in the coordinate representation. Comparing the result
with the generalized uncertainty relation, we derive the relation of the
thermal fluctuation and the information-entropies. From the results of this
paper, we find that the thermal fluctuation and quantum fluctuation are
separated.

\section{\textbf{\ RINDLER AND MINKOWSKI SPACE-TIME}}

\medskip

The coordinates in Rindler space-time can be obtained from the Minkowski
coordinates T, X under the transformations

\begin{eqnarray}
T &=&g^{-1}e^{g\xi }\sinh g\eta  \nonumber \\
X &=&g^{-1}e^{g\xi }\cosh g\eta \text{ \qquad \qquad for region R}
\end{eqnarray}
and

\begin{eqnarray}
T &=&-g^{-1}e^{g\widetilde{\xi }}\sinh g\widetilde{\eta }  \nonumber \\
X &=&-g^{-1}e^{g\widetilde{\xi }}\cosh g\widetilde{\eta }\text{ \qquad \quad
for region L}
\end{eqnarray}
where $g=$ constant $>$ $0,$ the Rindler coordinates ($\eta ,\xi $ ) and ($%
\widetilde{\eta },\widetilde{\xi }$ ) cover the space-time regions R and L
respectively. Both regions R and L are quadrants of Minkowski space-time as
shown in Fig.1. The region L is called the mirror space-time region of R.
With the method of standard Rindler quantization$^{13}$, we can obtain two
groups of annihilation and creation operators ($b$,$b^{\dagger }$) and ($%
\widetilde{b}$,$\widetilde{b}^{\dagger }$) corresponding to the Rindler
modes in the regions R and L, respectively, and the vacuum state defined by
these two groups of annihilation and creation operators is $\left|
0\right\rangle _R$ in region R and $\left| \widetilde{0}\right\rangle _R$ in
region L, respectively. The modes that are corresponding to these two groups
of annihilation and creation operators are complete in the region R and L
respectively, but they are not complete in the whole Minkowski space-time
region. Here the position is denoted by $\xi $ in region R, and by $%
\widetilde{\xi }$ in region L, while the momentum is denoted by $p_{_R}$in
region R, and by $\widetilde{p}_{_R}$ in region L.

The Minkowski vacuum is defined by general annihilation and creation
operators $\left( a,a^{\dagger }\right) $, $a\left| 0\right\rangle _M=0$.
Because of the different selecting of modes, we have another group of
Minkowski annihilation and creation operators $\left( d,d^{\dagger }\right) $
and $\left( \widetilde{d},\widetilde{d}^{\dagger }\right) $. The relations
between them and the Rindler annihilation operators satisfy the Bogoliubov
transformations

\begin{eqnarray}
d &\equiv &R(\theta )bR^{\dagger }(\theta )  \nonumber \\
\widetilde{d} &\equiv &R(\theta )\widetilde{b}R^{\dagger }(\theta )
\end{eqnarray}
where $\left[ d,d^{\dagger }\right] =\left[ \widetilde{d},\widetilde{d}%
^{\dagger }\right] =1$ . The unitary transformation (called thermal
transformation) is

\begin{equation}
R(\theta )=\exp \left\{ -\theta \left( \beta \right) \left( b\widetilde{b}%
-b^{\dagger }\widetilde{b}^{\dagger }\right) \right\}
\end{equation}
where

\begin{equation}
\tanh \left[ \theta \left( \beta \right) \right] =\exp \left( -\frac{\beta
\hbar \omega }2\right)
\end{equation}
$\beta =\frac 1{K_BT}$ , $K_B$ is the Boltzmann constant and $T$ is
temperature here. The vacuum state defined by$\left( d,d^{\dagger }\right) $
and $\left( \widetilde{d},\widetilde{d}^{\dagger }\right) $ is equivalent to
the Minkowski vacuum state $\left| 0\right\rangle _M$

\begin{equation}
d\left| 0\right\rangle _M=\widetilde{d}\left| 0\right\rangle _M=0
\end{equation}

For one-dimensional Rindler oscillator, we construct position and momentum $%
\left( x,p\right) $ from $\left( d,d^{\dagger }\right) $ and their tilde
conjugate quantities $\left( \widetilde{x},\widetilde{p}\right) $ from $%
\left( \widetilde{d},\widetilde{d}^{\dagger }\right) $ as follows:

\begin{eqnarray}
x &=&\sqrt{\frac \hbar {2m\omega }}\left( d+d^{\dagger }\right)  \nonumber \\
p &=&-i\sqrt{\frac{m\omega \hbar }2}\left( d-d^{\dagger }\right)
\end{eqnarray}
and

\begin{eqnarray}
\widetilde{x} &=&\sqrt{\frac \hbar {2m\omega }}\left( \widetilde{d}+%
\widetilde{d}^{\dagger }\right)  \nonumber \\
\widetilde{p} &=&-i\sqrt{\frac{m\omega \hbar }2}\left( \widetilde{d}-%
\widetilde{d}^{\dagger }\right)
\end{eqnarray}
The relation between Rindler vacuum and Minkowski vacuum is

\begin{equation}
\left| 0\right\rangle _M=R(\theta )\left| 0,\widetilde{0}\right\rangle _R%
\text{ ,}
\end{equation}
Where $\left| 0,\widetilde{0}\right\rangle _R$ is a direct product of the
Rindler vacuum state in region R and L, and $R(\theta )$ describes the
effect of a thermal bath in which a quantum harmonic oscillator immerses.
From Eq.(9), we can say loosely that a thermalizing operator $R(\theta )$
heats the ground state of a zero-temperature harmonic oscillator (Rindler
vacuum) into a thermal state with a finite temperature for a Rindler
uniformly accelerating observer. Note that any operator in region L commutes
with any tilde operator in region R for bosons in this paper. Consequently
any Minkowski vacuum expectation for the Rindler observer coincides with its
canonical ensemble average in statistical mechanics.

\medskip

\smallskip

\section{\textbf{\ INFORMATION-ENTROPY OF ONE-DIMENSIONAL RINDLER OSCILLATOR}
}

\smallskip

In this section, we will derive the reduced probability densities, and
discuss the information-entropies \smallskip associated with the
measurements of position and momentum for one-dimensional Rindler
oscillator. Comparing the generalized uncertainty relation of
one-dimensional Rindler oscillator, we present the relation of
information-entropies with quantum and thermal fluctuations. In this paper,
we use the information-entropy given by Shannon $^{14}$

\begin{equation}
S_A\left[ \Psi \right] =-\stackunder{\alpha }{\Sigma }\left| \left\langle
\alpha |\Psi \right\rangle \right| ^2\ln \left| \left\langle \alpha |\Psi
\right\rangle \right| ^2
\end{equation}
where $\left\{ \left| \alpha \right\rangle \right\} $ is the set of
eigenstates of $A$.

For the one-dimensional oscillator in Rindler space-time region R, its
Hamiltonian is

\begin{equation}
H=\frac 1{2m}p_R^2+\frac 12m\omega ^2\xi ^2=\left( b^{\dagger }b+\frac
12\right) \hbar \omega
\end{equation}
The wave function of ground state in the coordinate representation is

\begin{equation}
\left\langle \xi \mid 0\right\rangle _R=\left( \frac{m\omega }{\pi \hbar }%
\right) ^{\frac 14}\exp \left\{ -\frac{m\omega }{2\hbar }\xi ^2\right\}
\end{equation}
where , $p_{_R}=-i\hbar \frac d{d\xi }\equiv -i\hbar \partial _{_\xi },$ $m$
is the mass, $\omega $ is the angular frequency, and

\begin{eqnarray}
b &=&\frac 1{\sqrt{2m\hbar w}}\left( ip_{_R}+m\omega \xi \right)  \nonumber
\\
b^{\dagger } &=&\frac 1{\sqrt{2m\hbar \omega }}\left( -ip_{_R}+m\omega \xi
\right)
\end{eqnarray}
are the corresponding annihilation and creation operators in the Rindler
region R, respectively. Using the tilde rules in Thermal field dynamics, we
introduce

\begin{equation}
\widetilde{H}=\frac 1{2m}\widetilde{p}_{_R}^2+\frac 12m\omega ^2\widetilde{%
\xi }^2=\left( \widetilde{b}^{\dagger }\widetilde{b}+\frac 12\right) \hbar
\omega
\end{equation}
in Rindler region L. Substituting Eq.(13) into Eq.(4), one has

\begin{equation}
R(\theta )=\exp \left\{ i\frac \theta \hbar \left( \xi \widetilde{p}_{_R}-%
\widetilde{\xi }p_{_R}\right) \right\}
\end{equation}
with $\theta \equiv \theta \left( \beta \right) $ . From Appendix B.4 in
Ref.15, the last formula can be written as

\begin{eqnarray}
R(\theta ) &=&\exp \left\{ -\tanh \left( \theta \right) \widetilde{\xi }%
\partial _{_\xi }\right\} \exp \left\{ \ln \left[ \cosh \left( \theta
\right) \right] \left( \xi \partial _{_\xi }-\widetilde{\xi }\partial _{%
\widetilde{\xi }}\right) \right\}  \nonumber \\
&&\cdot \exp \left\{ -\tanh \left( \theta \right) \xi \partial _{\widetilde{%
\xi }}\right\} \text{ .}
\end{eqnarray}
Using the following operator properties

\begin{equation}
e^{C\partial _y}f\left( y\right) =f\left( y+C\right)
\end{equation}
and

\begin{equation}
e^{Cy\partial _y}f\left( y\right) =f\left( ye^C\right) \text{ ,}
\end{equation}
one can gave the wave function of Minkowski vacuum in Rindler coordinate
representation

\begin{eqnarray}
&&\left\langle \widetilde{\xi },\xi \mid 0\right\rangle _M  \nonumber \\
&=&R(\theta )\left( \frac{m\omega }{\pi \hbar }\right) ^{\frac 12}\exp
\left\{ -\frac{m\omega }{2\hbar }\left( \xi ^2+\widetilde{\xi }^2\right)
\right\}  \nonumber \\
&=&\left( \frac{m\omega }{\pi \hbar }\right) ^{\frac 12}\exp \{-\frac{%
m\omega }{2\hbar }[\left( \xi \cosh \left( \theta \right) -\widetilde{\xi }%
\sinh \left( \theta \right) \right) ^2  \nonumber \\
&&+\left( \widetilde{\xi }\cosh \left( \theta \right) -\xi \sinh \left(
\theta \right) \right) ^2]\}.
\end{eqnarray}

We can see that when $\beta \rightarrow \infty $, from Eq.(4) and Eq.(5) we
have $\theta \left( \beta \right) \rightarrow 0,$ $R(\theta )\rightarrow 1,$
so $\left\langle \widetilde{\xi },\xi \mid 0\right\rangle _M$ is reduced to $%
\left\langle \widetilde{\xi },\xi \mid 0,\widetilde{0}\right\rangle _R.$

The matrix element of density operator in position presentation for
one-dimensional Rindler oscillator is

\begin{eqnarray}
\rho _{_{\xi ^{\prime },\xi }} &=&\int_{-\infty }^{+\infty }\left\langle \xi
^{^{\prime }},\widetilde{\xi }|0\left( \beta \right) \right\rangle
\left\langle 0\left( \beta \right) |\widetilde{\xi },\xi \right\rangle d%
\widetilde{\xi }  \nonumber \\
&=&\frac{m\omega }{\pi \hbar }\int_{-\infty }^{+\infty }\exp \left\{ -\frac{%
m\omega }{2\hbar }\left[ \left( \xi ^{^{\prime }}\cosh \left( \theta \right)
-\widetilde{\xi }\sinh \left( \theta \right) \right) ^2+\left( \widetilde{%
\xi }\cosh \left( \theta \right) -\xi ^{^{\prime }}\sinh \left( \theta
\right) \right) ^2\right] \right\}  \nonumber \\
&&\cdot \exp \left\{ -\frac{m\omega }{2\hbar }\left[ \left( \xi \cosh \left(
\theta \right) -\widetilde{\xi }\sinh \left( \theta \right) \right)
^2+\left( \widetilde{\xi }\cosh \left( \theta \right) -\xi \sinh \left(
\theta \right) \right) ^2\right] \right\} d\widetilde{\xi }  \nonumber \\
&=&\sqrt{\frac{m\omega }{\pi \hbar \left( \cosh ^2\left( \theta \right)
+\sinh ^2\left( \theta \right) \right) }}\cdot \exp \{-\frac{m\omega }{%
2\hbar }[\left( \cosh ^2\left( \theta \right) +\sinh ^2\left( \theta \right)
\right) \left( \xi ^{^{\prime }2}+\xi ^2\right)  \nonumber \\
&&-\frac{2\sinh ^2\left( \theta \right) \cosh ^2\left( \theta \right) \left(
\xi +\xi ^{^{\prime }}\right) ^2}{\cosh ^2\left( \theta \right) +\sinh
^2\left( \theta \right) }]\}
\end{eqnarray}
Taking $\xi =\xi ^{^{\prime }}$ in Eq.(20), one has the probability density
of position

\begin{eqnarray}
\rho _{_{\xi ,\xi }} &=&\sqrt{\frac{m\omega }{\pi \hbar \left( \cosh
^2\left( \theta \right) +\sinh ^2\left( \theta \right) \right) }}\exp
\left\{ -\frac{m\omega }\hbar \cdot \frac 1{\cosh ^2\left( \theta \right)
+\sinh ^2\left( \theta \right) }\xi ^2\right\}  \nonumber \\
&=&\sqrt{\frac{m\omega }{\pi \hbar \cosh \left( 2\theta \right) }}\exp
\left\{ -\frac{m\omega }\hbar \cdot \frac 1{\cosh \left( 2\theta \right)
}\xi ^2\right\}
\end{eqnarray}

Similarly, we have the reduced probability density of momentum for
one-dimensional Rindler oscillator

\begin{eqnarray}
\rho _{_{p_R,p_R}} &=&\int_{-\infty }^{+\infty }\frac 1{2\pi \hbar }\exp
\left\{ i\frac{p_R\xi ^{^{\prime }}}\hbar -i\frac{p_R\xi }\hbar \right\}
\rho _{\xi ^{^{\prime }},\xi }d\xi d\xi ^{^{\prime }}  \nonumber \\
&=&\sqrt{\frac 1{m\omega \pi \hbar \cosh \left( 2\theta \right) }}\exp
\left\{ -\frac 1{m\omega \hbar }\cdot \frac 1{\cosh \left( 2\theta \right)
}p_R^2\right\}
\end{eqnarray}
So, we can calculate the information-entropy with the measurements of
position and momentum, respectively

\begin{eqnarray}
s_{_{_\xi }} &=&-\int_{-\infty }^{+\infty }\rho _{_{\xi ,\xi }}\ln \left(
\rho _{_{\xi ,\xi }}\right) d\xi  \nonumber \\
&=&-\int_{-\infty }^{+\infty }\sqrt{\frac{m\omega }{\pi \hbar \cosh \left(
2\theta \right) }}\exp \left\{ -\frac{m\omega }\hbar \cdot \frac 1{\cosh
\left( 2\theta \right) }\xi ^2\right\}  \nonumber \\
&&\left[ \left( -\frac{m\omega }\hbar \cdot \frac 1{\cosh \left( 2\theta
\right) }\xi ^2\right) +\ln \left( \sqrt{\frac{m\omega }{\pi \hbar \cosh
\left( 2\theta \right) }}\right) \right] d\xi  \nonumber \\
&=&\frac 12\left[ 1+\ln \pi +\ln \left( \cosh \left( 2\theta \right) \right)
+\ln \left( \frac \hbar {m\omega }\right) \right]
\end{eqnarray}
and

\begin{eqnarray}
s_{_{p_R}} &=&-\int_{-\infty }^{+\infty }\rho _{_{p_R,p_R}}\ln \left( \rho
_{_{p_R,p_R}}\right) dp_R  \nonumber \\
&=&-\int_{-\infty }^{+\infty }\sqrt{\frac 1{m\omega \pi \hbar \cosh \left(
2\theta \right) }}\exp \left\{ -\frac 1{m\omega \hbar }\cdot \frac 1{\cosh
\left( 2\theta \right) }p_R^2\right\}  \nonumber \\
&&\left[ \left( -\frac 1{m\omega \hbar }\cdot \frac 1{\cosh \left( 2\theta
\right) }p^2\right) +\ln \left( \sqrt{\frac 1{m\omega \pi \hbar \cosh \left(
2\theta \right) }}\right) \right] dp_R  \nonumber \\
&=&\frac 12\left[ 1+\ln \pi +\ln \left( \cosh \left( 2\theta \right) \right)
+\ln \left( m\omega \hbar \right) \right]
\end{eqnarray}

Deutsch and Partovi$^{1,2}$discussed the sum of entropies associated with
the measurements of a generic non-commutative pair of observable (A,B) in a
normalized state $\left| \Psi \right\rangle $

\begin{equation}
\stackunder{}{U\left[ A,B:\psi \right] =S_A\left[ \psi \right] +S_B\left[
\psi \right] }\text{ ,}
\end{equation}
which cannot be made arbitrarily small but has an irreducible lower bound
independent of the choice of $\left| \Psi \right\rangle $ .

Thus, we get the sum of information-entropies \smallskip associated with the
measurements of position and momentum of one-dimensional Rindler oscillator.

\begin{equation}
U=s_{_{_\xi }}+s_{p_R}=1+\ln \pi +\ln \left( \cosh \left( 2\theta \right)
\right) +\ln \hbar
\end{equation}
where $U$ denotes the uncertainty measurement of one-dimensional Rindler
oscillator.

Now we wish to find the relation of information-entropy and fluctuation. In
our previous paper$^4$, according to the invariance of Bogoliubov
transformation in the Thermal Field Theory, we derived the generalized
uncertainty relation of one-dimensional Rindler oscillator in Minkowski
vacuum in the coordinates representation, that is

\begin{equation}
\left\langle \left( \Delta p_R\right) ^2\right\rangle \left\langle \left(
\Delta \xi \right) ^2\right\rangle \geq \frac{\hbar ^2}4+\frac{\hbar ^2}{%
4\sinh ^2\left( \frac{\beta \omega \hbar }2\right) }\text{ .}
\end{equation}
For a Rindler uniformly accelerated observer, the term on the LHS(the left
hand side )of Eq.(27) describes the total fluctuations of one-dimensional
Rindler oscillator. The first term on the RHS of Eq.(27) $\left\langle
\left( \Delta p\right) ^2\right\rangle \left\langle \left( \Delta x\right)
^2\right\rangle $ describes the zero-temperature fluctuation, which is a
purely quantum fluctuation and satisfies the general uncertainty relation.
The second term on the RHS of Eq.(27) describes a purely thermal fluctuation
of one-dimensional Rindler oscillator, which is determined by cross terms of
the tilde and non-tilde operators.

The thermal fluctuation can be written as

\begin{equation}
\frac{\hbar ^2}{4\sinh ^2\left( \frac{\beta \omega \hbar }2\right) }=\frac{%
\hbar ^2}4\left( \cosh ^2\left( 2\theta \right) -1\right)
\end{equation}
Where $\theta $ defined by Eq.(5). Comparing Eq.(26) with Eq.(28), one has

\begin{equation}
\frac{\hbar ^2}{4\sinh ^2\left( \frac{\beta \omega \hbar }2\right) }=\frac
14\hbar ^2\left( \frac{e^{2\left( s_{_{_\xi }}+s_{p_{_R}}\right) }}{e^2\pi
^2\hbar ^2}-1\right)
\end{equation}

Hence, we derive the relation of information-entropy, quantum fluctuation
and thermal fluctuation

\begin{equation}
\frac{e^{2\left( s_{_{_\xi }}+s_{p_{_R}}\right) }}{4e^2\pi ^2}=\frac{e^{2U}}{%
4e^2\pi ^2}=\frac 14\hbar ^2+\frac{\hbar ^2}{4\sinh ^2\left( \frac{\beta
\omega \hbar }2\right) }
\end{equation}

\section{\textbf{\ SUMMARY AND DISCUSSION}}

\smallskip

In Eq.(26) with $\hbar =1$, when temperature $T\rightarrow 0$ , we get

\begin{equation}
S_{_\xi }+S_{P_R}=1+\ln \pi
\end{equation}
This result coincides with the results obtained by other methods$^{16}$.

The Eq.(30) is the key result of this paper. The term on the LHS of the
Eq.(30) includes the sum of information-entropies \smallskip associated with
the measurements of position and momentum of one-dimensional Rindler
oscillator. The first term on the RHS of Eq.(30) $\frac 14\hbar ^2$
describes zero-temperature fluctuation, which is the purely quantum
fluctuation. The second term on the RHS of Eq.(30) describes the purely
thermal fluctuation of one-dimensional Rindler oscillator. Thus we can make
the thermal fluctuation and quantum fluctuation separated naturally. When
temperature increases, $S_{_\xi }+S_{P_R}$ also increases monotonously. This
result shows that the thermal fluctuation causes the loss of information.

According to Eq.(27) and (30), one have 
\begin{equation}
\left\langle \left( \Delta p_R\right) ^2\right\rangle \left\langle \left(
\Delta \xi \right) ^2\right\rangle \geq \frac{e^{2\left( s_{_{_\xi
}}+s_{p_R}\right) }}{4e^2\pi ^2}
\end{equation}
This is the relation between the uncertainty and the information-entropy.
When temperature $T\rightarrow 0,$ the quantum uncertainty relation will be
restored.


\smallskip

\begin{thebibliography}{99}
\bibitem{}  D.Deutcsh, \textit{Phys.Rev.Lett.} \textbf{50}, 631 (1983).

\bibitem{}  M.H.Patrovi, \textit{Phys.Rev.Lett}. \textbf{50}, 1883 (1983).

\bibitem{}  C.E.Shannon and W.W.Weaver, \textit{Mathematical Theory of
Communication }(University Illinois Press, 1949).

\bibitem{}  X.Ye and Y.X.Gui, \textit{Int}. \textit{J. Theo.} \textit{Phys.}%
, \textbf{40, }1341(2001)

\bibitem{}  A. Kempf, G. Mangano, R. B. Mann \textit{Phys. Rev., }D 52, 1108
(1995)

\bibitem{}  D.V.Ahluwalia, \textit{Phys. Lett.} A \textbf{275} 31(2000)

\bibitem{}  S.de Haro, \textit{Class. Quantum. Grav.} \textbf{15} 519 (1998)

\bibitem{}  A.Mann, M.Revzen, H.Umezawa and Yamanaka, \textit{Phys.Lett}.%
\textbf{\ A 140} 475 (1989)

\bibitem{}  W.G.Unruh, \textit{Phys.Rev.}D , \textbf{14}, 870 (1976).

\bibitem{}  J. Hughes Richard, \textit{Ann. Phys.}, \textbf{162, }1 (1985)

\bibitem{}  C.Anstropoulos and J.J.Halliwell, \textit{gr-qc/}9407039

\bibitem{}  A.Anderson and J.J.Halliwell, \textit{Phys. Rev.}, \textbf{D 48}%
, 2753(1993)

\bibitem{}  N.D.Birrell, and P.C.W. Davies, \textit{Quantum Fields in Curved
Space} (Cambridge University Press, 1982).

\bibitem{}  S.Abe and N.Suzuki, \textit{Phys.Rev.} \textbf{A 41}, 4608(1990)

\bibitem{}  D.A.Kirzhnits, \textit{Filed Theoretical Methods in Many-body
Systems} (Pergamon Press, 1967)

\bibitem{}  H.Evertt III, \textit{The many-worlds interpretation of quantum
Mechanics }(Princeton University Press, 1973).
\end{thebibliography}
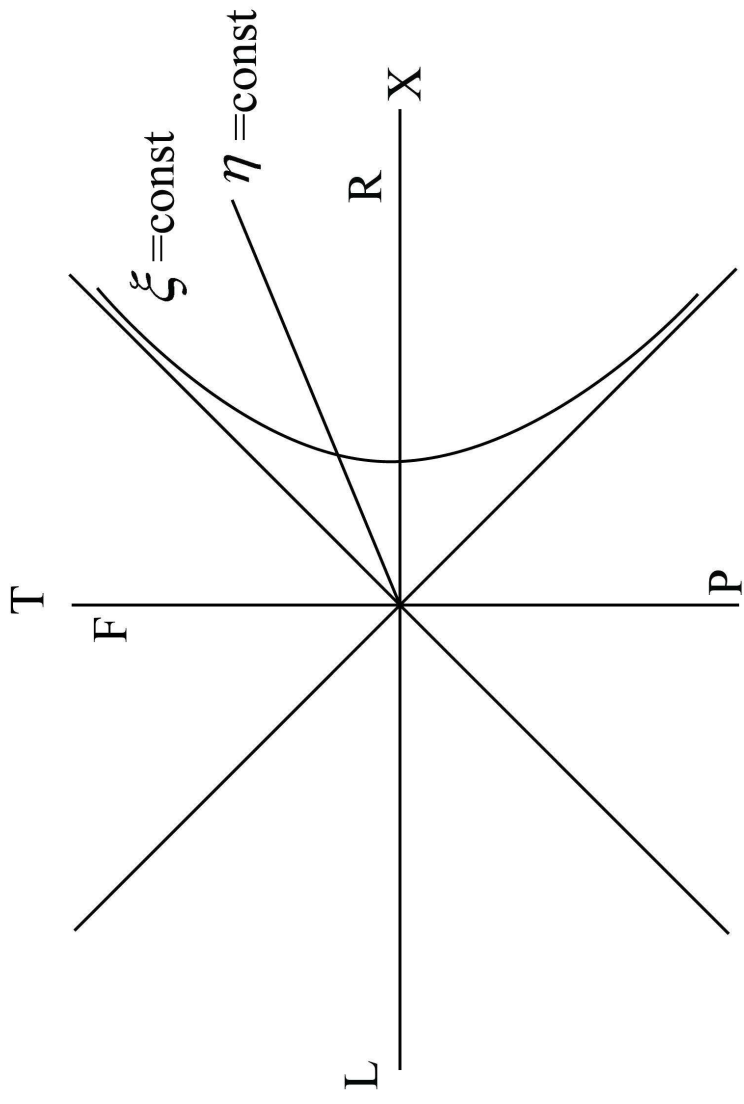
\end{document}